\documentclass[reprint,amsmath,amssymb,aps,prl]{revtex4-1}

\usepackage{graphicx}
\usepackage{dcolumn}
\usepackage{bm}
\usepackage{hyperref}

\begin{document}

\preprint{APS/123-QED}

\title{Janus Points and Arrows of Time}

\author{Julian Barbour}
\affiliation{College Farm, South Newington, Banbury, Oxon, OX15 4JG UK,\\
Visiting Professor in Physics at the University of Oxford, UK.}

\author{Tim A. Koslowski}
\affiliation{Universidad Nacional Aut\'onoma de M\'exico - Instituto de Ciencias Nucleares\\
 Apdo. Postal 70-543, 04510 M\'exico, DF, M\'exico
}

\author{Flavio Mercati}
\affiliation{Perimeter Institute for Theoretical Physics,\\ 
31 Caroline Street North, Waterloo, ON, N2L 2Y5 Canada.}

\begin{abstract}
  \noindent We clarify and strengthen our demonstration that arrows of time necessarily arise in unconfined systems. Contrary to a recent claim, this does not require an improbable selection principle.
\end{abstract}

\maketitle

{\bf Introduction.} In \cite{zeh}, Zeh criticizes our \cite{PRL,AoT,Entaxy}. In response, we first review our papers. We consider $N$-body model universes to see if time-reversal symmetric dynamics can lead, without special conditions, to evolutions that inhabitants of the universe will experience as having a direction of time. Our systems are \emph{unconfined} and, like the universe, can expand freely. In contrast, all of thermodynamics and statistical mechanics presupposes systems that are confined \cite{barb}, either by man or nature: their phase space of accessible states is bounded. We point out that unconfined systems require \emph{different conceptualization}.

First, all measurements are ratios of an interval and a reference unit. In confined systems, time is defined by an external clock and scale by the `box' that confines it. For the universe, instead, one needs to take a solution of the model and define physical units using processes that unfold within it.

Second, external structures like a preferred inertial frame of reference must not play a role. This is so for $N$-body solutions for which the energy $E$, linear momentum ${\bf P}$ and angular momentum ${\bf L}$ all vanish. Such solutions constitute \emph{the relational $N$-body problem}. It is defined by shape degrees of freedom, the coordinates of \emph{shape space} ${\sf S}$: angles and dimensionless ratios of inter-particle distances. All objective predictions must be deduced from the unparametrized and undirected solution curves in ${\sf S}$. This is \emph{shape dynamics} \cite{GGK} \cite{tutorial}. It describes the universe in intrinsic terms. The key quantity in \cite{PRL,AoT,Entaxy} is the \emph{complexity} $C$, a scale-invariant function on {\sf S}:
\begin{equation}
C=-{1\over m_\textrm{tot}^{5/2}}\sqrt{I_\textrm{cm}}\,V_\textrm{New},\label{com}
\end{equation} 
in which $I_\textrm{cm}$ is the centre-of-mass moment of inertia:
\begin{equation}
I_\textrm{cm}=\sum_{a=1}^Nm_a{\bf r}_a^\textrm{cm}\cdot{\bf r}_a^\textrm{cm}\equiv \sum_{a<b}{m_am_b\over m_\textrm{\,tot}}\,r_{ab}^2:=m_\textrm{tot}\ell_\textrm{rms}^2,\label{rms}
\end{equation}
and $V_\textrm{New}$ is the Newton potential:
\begin{equation}
-V_\textrm{New}=\sum_{a<b}{m_am_b\over r_{ab}}:=m_\textrm{tot}^2\ell_\textrm{mhl}^{-1}.\label{new}
\end{equation}
In (\ref{rms}), ${\bf r}_a^\textrm{cm}$ is the position of particle $a$ relative to the centre of mass, $m_\textrm{\,tot}=\sum_am_a$ and $r_{ab}=|{\bf r}_a-{\bf r}_b|$.
We shall see that the root-mean-square length $\ell_\textrm{rms}$ (\ref{rms}) and mean harmonic length $\ell_\textrm{mhl}$ (\ref{new}) play an interesting role.

The complexity $C$ is a sensitive measure of clustering since the centre-of-mass moment of inertia $I_\textrm{cm}$ (twice the inertia tensor's trace) changes little if particles cluster whereas the Newton potential $V_\textrm{New}$ grows sharply. The behaviour of $C$, an observable on {\sf S}, indicates whether the evolution exhibits an arrow of time. Remarkably, $C$ also controls the dynamics: the forces that change the universe's shape derive from it. These are the only forces inhabitants of the model universe can observe.

The fundamental difference between confined and unconfined systems is this. In confined systems on which no external forces  act (except those of the confining `box'), Poincar\'e recurrence is generally realized. Except for tiny (and, very rarely, large) fluctuations, the Boltzmann entropy $S_\textrm B$ is nearly always at the thermal-equilibrium maximum. The solutions are qualitatively time-reversal symmetric.

In unconfined systems the solution structure is very different. In \cite{PRL, AoT, Entaxy}, we consider the relational $N$-body problem and, in \cite{Entaxy}, pure inertial motion. In both cases, $I_\textrm{cm}$ has a unique minumum that (except for a zero-measure set) divides all solutions in half. We call this the \emph{Janus point} $J$: the two halves exhibit emergent arrows of time that, like the Roman god, point in opposite directions away from $J$. This is clearly manifested in the relational $N$-body problem, in which $C$ always has a minimum near $J$ and grows, with fluctuations but between rising bounds, in both directions away from $J$. 

The growth of $C$ is due to bound-cluster formation; as noted, clustering causes $C$ to grow. Of course, gravity has long been seen as `anti-thermodynamic' because it has the opposite effect to the uniformization intuitively associated with entropy growth. However, prior to our \cite{AoT, PRL}, it seems to have escaped notice that, by the time-reversal symmetry of the $N$-body problem (with $E\ge 0$), clustering must also occur in the opposite time direction with a minimum of $C$ somewhere in the middle. Together with the recognition that all objective information is expressed through shape degrees of freedom, this was a first hint that the various arrows of time could have a dynamical origin unrelated to a special selection condition. It already showed this for the arrow associated with the formation of structure and records.

Further unexpected encouragement came when we realized the significance \cite{Entaxy} of unbounded growth in unconfined systems of the scale degree of freedom ($I_\textrm{cm}$ in the $N$-body problem or the volume of the universe in general relativity). The point is that, in the phase space with scale, Liouville's theorem enforces conservation of phase-space volume. If the scale degree of freedom grows, so too must the volume of the scale part of the phase space occupied by a Gibbs ensemble. This means that the shape part must get smaller. \emph{The shape degrees of freedom will be subject to attractors.} 

This alone reveals a deep dynamical origin of several arrows of time: the formation of structure and records in the universe, the second law of thermodynamics in local subsystems of the universe (as outlined in \cite{Entaxy}) and the outgoing waves generated by retarded potentials. The quantum-mechanical arrow of time (collapse of the wave function) might also have the same origin. That growth of the scale phase-space volume must decrease the remaining part is also noted (in connection with cosmological inflation) in \cite{sloan}.

Having reviewed \cite{PRL,AoT,Entaxy} we now respond to \cite{zeh}.

{\bf Response to Zeh.} Zeh states:

1.~\emph{The philosophy of shape dynamics is wrong if it requires that scale be physically meaningless.} We do not deny scale's role in the extended phase space in which Liouville's theorem holds. We assert that only quantities defined in shape space are observable. Nature as we experience it locally with our unit conventions is not scale invariant. Moreover, our model-universe solutions contain stable structures within them that provide internally generated standards of length and time in the form of the major axes and periods of Kepler pairs. All physical observations are expressed in terms of ratios and are therefore scale-invariant. The global scale is an unobservable auxiliary concept that permits a phase-space conserving Hamiltonian description of the universe's evolution. The phase space of the \emph{observable} shape degrees of freedom is not conserved.

2.~{\it Rescaling of time is incompatible with Newtonian mechanics.} Newton's equations are not time-rescaling invariant but they exhibit dynamical similarity: the simultaneous rescaling of time $t\to \lambda\, t$, positions $ {\bf r}_i \to \lambda^{2/3}\, {\bf r}_i$ and momenta ${\bf p}_i \to \lambda^{-1/3}\, {\bf p}_i$ maps one Newtonian solution into another. Two such solutions coincide when projected to unparametrized curves in shape space. They are indistinguishable for the hypothetical inhabitants of our model universe and physically equivalent. Dynamical similarity is an essential element in the shape-dynamical description of conventional physics.

3.~{\it  The quantum-mechanical $\hbar$ is a fundamental unit.} It did eliminate the undetermined additive constant in classical entropy, making it possible to count quantum states and associate the phase-space volume $\hbar$ with each. But quantum states require boundary conditions, which presuppose some kind of confinement \cite{barb}, so Zeh's argument is not decisive in the situations we consider. In fact, in \cite{AoT} we tentatively proposed a scale-invariant wave equation of an unconfined universe. It contains no dimensionful constant. We expect a dimensionful $\hbar$ to emerge in the universe's subsystems. If so, the apparently fundamental $\hbar$ scale will admit a precise formulation in shape space: all atoms have identical shapes and have the same size when next to each other. The ontological primacy of shape space will be maintained.

4.~{\it Solutions with a Janus point require an improbable condition to be applied at the start of the calculation}. This misunderstanding may have arisen because we give examples of numerical solutions, which we could only obtain by specifying initial conditions. However, the fundamental fact, well known in $N$-body theory, is that in \emph{all} (apart from a zero-measure set) solutions of our model $I_\textrm{cm}$ has a unique minimum, the Janus point. This is a direct consequence \cite{PRL} of Newton's laws, the homogeneity of degree $-1$ of the Newton potential and its non-negativity. Whatever the initial conditions of a numerical solution, it must have a unique Janus point somewhere. 

5.~{\it Our Janus-point condition is improbable because a generic distribution of a finite number of objects in infinite space would never be found in a finite volume}. This is incorrect for the reason just given, but it raises a fundamental point. As Gibbs \cite{gibbs} emphasizes, statistical arguments can only be applied to normalizable probability distributions. This requires particles to be confined to a finite spatial volume.  Then, given $N$ particles in a volume $V$, the probability of finding them by chance in a volume $v\subset V, v\ll V,$ tends to zero in the limit $v/V\rightarrow 0$. This, we think, is Zeh's intuition. But it is no help without the extrinsic scale $V$. Therefore, we need to characterize distributions of particles in an unconfined space \emph{intrinsically} by means of ratios. 

This can be done. Mathematics supplies two mass-weighted lengths perfectly suited to this end: $\ell_\textrm{rms}$ (\ref{rms}) and $\ell_\textrm{mhl}$ (\ref{new}), which reflect the greatest and least inter-particle separations, respectively. Without an extrinsic scale, these lengths by themselves have no meaning, but their ratio $C$ (\ref{com}) does. It measures the extent to which the particles are clustered. We see no alternative to the use of such scale-invariant quantities in the creation of a statistical theory of unconfined systems \cite{Entaxy}. This is why the degree to which points are clustered is important: it is independent of scale. 

In any solution of the relational $N$-body problem $C$ is epoch dependent and always has a minimum near the point at which, in the configuration space with scale, $I_\textrm{cm}$ has its minimum. Either side of its minimum, $C$ grows with fluctuations between monotonically rising bounds. This Janus-type behaviour reveals the objective presence of oppositely pointing arrows of experienced time. 

6.~{\it Our $N$-body Janus-point structure also exists for free inertial particles and this is trivial}. This is true but still important (the proof of Russell's paradox shows that just because the proof of a theorem is trivial that does not mean the theorem is trivial). Indeed, this simplest of all dynamical systems (purely inertial motion) has not alerted researchers to the difference between confined and unconfined systems. We also note that Zeh attributes a \emph{minimum} of the entropy to the system when at its Janus point and concludes our gravitational arrow is just the thermodynamic arrow in disguise. But, as we emphasize in \cite{Entaxy},~Sec.~2, and \cite{barb}, it is only near the Janus point that the state is disorderly; away from it, the attractors force the motion into very special Hubble-type expansion. This hardly suggests an entropy minimum at the Janus point. Indeed, as we argue in \cite{Entaxy}, any entropy-type quantity defined for an unconfined universe in its totality must \emph{decrease} (not increase) in both directions away from the Janus point. 

Our conclusion for the universe can be reconciled with the second law of thermodynamics, the validity of which has been established \emph{only for subsystems} of the universe -- and moreover subject to the strictly controlled conditions described in \cite{barb}. A purely Newtonian gravitational derivation of the second law is outlined in \cite{Entaxy}: the inevitable shape attractors in a self-gravitating expanding universe lead to clustering and the creation of quasi-bound subsystems. Being effectively self-confined, one can define for them a Boltzmann entropy that increases in almost all cases in the same direction as the universe becomes more structured and its entaxy decreases. 

If a realistic law of the universe based on general relativity with appropriate matter can be shown to enforce Janus-type solutions, worries about the behaviour of entropy and the origin to the various arrows of time may turn out to be an artefact of an implicitly assumed but non-existent `container of the universe'.

\end{document}